\documentclass[twocolumn,pra,showpacs,aps]{revtex4-1}

\usepackage[english]{babel}
\usepackage[ansinew]{inputenc}
\usepackage{times}
\usepackage{graphicx}
\usepackage{graphics}
\usepackage{amsmath}
\usepackage{amsfonts}
\usepackage{amssymb}
\usepackage{dsfont}
\usepackage{epstopdf}
\usepackage{makeidx}
\usepackage{subfigure}
\usepackage{color}
\usepackage{pgf}
\usepackage{bm}
\usepackage{tikz} % Create PostScript and PDF graphics in TeX. Depends on the PGF package.
\usepackage[normalem]{ulem}

\newcommand{\be}{\begin{equation}}
\newcommand{\ee}{\end{equation}}
\newcommand{\bea}{\begin{eqnarray}}
\newcommand{\eea}{\end{eqnarray}}

\makeindex
\begin{document}
%\doublespacing
\title{Spontaneous emission in the presence of a spherical plasmonic cloak}
\author{W. J. M.  Kort-Kamp}
\affiliation{Instituto de F\'{\i}sica, Universidade Federal do Rio de Janeiro,
Caixa Postal 68528, Rio de Janeiro 21941-972, RJ, Brazil}
\author{F. S. S. Rosa}
\affiliation{Instituto de F\'{\i}sica, Universidade Federal do Rio de Janeiro,
Caixa Postal 68528, Rio de Janeiro 21941-972, RJ, Brazil}
\author{F. A. Pinheiro}
\affiliation{Instituto de F\'{\i}sica, Universidade Federal do Rio de Janeiro,
Caixa Postal 68528, Rio de Janeiro 21941-972, RJ, Brazil}
\author{C. Farina}
\affiliation{Instituto de F\'{\i}sica, Universidade Federal do Rio de Janeiro,
Caixa Postal 68528, Rio de Janeiro 21941-972, RJ, Brazil}

\date{\today}

\begin{abstract}

We investigate the spontaneous emission of a two-level atom placed in the vicinities of a plasmonic cloak composed of a coated sphere. In the dipole approximation, we show that the spontaneous emission rate can be reduced to its vacuum value provided the atomic emission frequency lies within the plasmonic cloak frequency operation range. Considering the current status of plasmonic cloaking devices, this condition may be fulfilled for many atomic species so that we argue that atoms with a sufficiently strong transition can be used as quantum, local probes for the efficiency of plasmonic cloaks.

\end{abstract}
\maketitle
%
% -------------------------------------------------------------------------------------------------
\section{Introduction}\label{intro}

Among the most interesting effects of Cavity Quantum Electrodynamics (Cavity QED) are the possible influences that bodies in the vicinities of atomic systems may exert in their radiative properties. The fact that the environment can strongly modify the spontaneous emission (SE) rate of a system was theoretically predicted for the first time (to the best of our knowledge) by Purcell in 1947, in a national conference of the American Physical Society \cite{Purcell-46}. Approximately two decades later, Drexhage {\it et al} \cite{DrexhageEtAl-68} observed the variation of the fluorescence decay time of a molecule near a mirror. Also, Morawitz \cite{Morawitz-69} obtained the SE rate of a two-level system near a perfectly conducting mirror and showed that the result exhibited oscillations with the distance between the system and the mirror, showing that the SE rate can be enhanced or suppressed. The first experiments that confirmed these spacial oscillations for were done again by Drexhage in 1970 \cite{Drexhage-70} and in 1974 \cite{Drexhage-74}. Generalizations for an atom between two parallel conducting plates were done by many authors \cite{Barton-70-87,Stehle-70,Milonni-Knight-73,Philpott-73,Alves-Farina-Tort-2000,MendesEtAl-08}, and the possibility of modifying the emission properties of atoms and molecules in a controlled way has stimulated the investigation of SE in different systems and geometries \cite{blanco2004,thomas2004,carminati2006,vladimirova2012,klimovreview}. Even in simple situations as those with an atom between two parallel mirrors, quite interesting phenomena may occur. For instance, if the transition dipole moment is parallel to the mirrors, the SE rate may be significantly suppressed if the distance between the plates is shortened below a critical value~\cite{Barton-70-87}. Suppression of SE rate was observed by the first time in 1985 by Hulet {\it et al} \cite{Hulet-Hilfer-Kleppner-85} by using a beam of excited Cesium atoms passing between mirrors and later on by other groups~\cite{JheEtAl-87,DeMartiniEtAl-87,HeinzenEtAl-87}. There are excellent reviews on how strongly the environment can modify the SE rate of an emitter (such as atoms, molecules, or quantum dots), see, for instance, Refs \cite{klimovreview,kuhnreview,Milonni,Haroche-Kleppner-89,Hinds-90,Berman-94}.

 In the last decade the research on the radiative properties of atoms and molecules has received renewed interest due to progress in near-field optics. Indeed, advances in nano-optics have not only allowed the increase of the spectroscopical resolution of molecules in complex environments~\cite{betzig1993} but have also led to the use of nanometric objects (e.g. nanoparticles and nano-tips) that to modify the lifetime and to enhance the fluorescence of single molecules~\cite{bian1995,sanchez1999}. Another important example is the development of nanoantennas that can enhance the local optical field and modify the fluorescence of single-emitters~\cite{greffet2005,muhlschlegel2005}.

The advent of plasmonic devices has also opened new possibilities for tailoring the emission properties of atoms and molecules. When a quantum emitter is located near a plasmonic structure it may experience an enhancement of the local field due the excitation of a plasmonic resonance, which affects the lifetime of an excited state. This effect has been exploited in the development of important applications in nanoplasmonics, such as surface-enhanced fluorescence and surface-enhanced Raman scattering~\cite{jackson2004,wei2008,li2010} and the modification of two-level atom resonance fluorescence~\cite{klimov2012}. It is also very important to point out the role that metamaterials may have on quantum emitters' radiative processes. Metamaterials are artificial structures with engineered electromagnetic response that may exhibit unusual properties such as negative refraction~\cite{Smith00}, resolution of images beyond the diffraction limit~\cite{smith2004}, optical magnetism~\cite{enkrich2005,cai2007}, electromagnetic cloaking~\cite{pendry2006,leonhardt2006} and support of slow light propagation~\cite{OPN}. Although the influence of some of these properties, such as negative refraction, on the SE of atoms has been investigated~\cite{klimov2002}, many others remain unexplored so far. Recently, nanostructured media with hyperbolic dispersion have emerged as a new class of metamaterials with many applications in biosensing, subwavelength imaging and waveguiding~\cite{hyperbolicreview}. In addition, hyperbolic metamaterials exhibit a broadband enhancement in the electromagnetic density of states that has been demonstrated to have an important impact on the SE of molecules and quantum dots, allowing for several applications in quantum nanophotonics~\cite{jacob2012,hyperbolicreview}.

The aim of the present paper is to investigate the emission properties of an atom in the presence of a cloaking device. We specifically examine the behavior of the SE decay rate of an atom near a plasmonic cloak. The concept of plasmonic cloaking, theoretically proposed by Al\`u and Engheta~\cite{alu2005}, is based on the scattering cancelation technique, in which a dielectric or conducting object can be effectively cloaked by covering it with a homogeneous and isotropic layer of plasmonic material with low-positive or negative electric permittivity. In these systems, the incident radiation induces a local polarization vector out-of-phase with respect to the local electric field so that the in-phase contribution given by the scattering object may be partially or totally canceled~\cite{alu2005,alu2008}. The first experimental realization of this idea was recently implemented for microwaves~\cite{edwards2009}, paving the way for many applications in camouflaging, low-noise measurements and non-invasive sensing. Here we focus on a novel application of plasmonic cloaking in atomic physics, demonstrating that the spontaneous emission rate of an atom can be drastically modified in the presence of a plasmonic cloak. To the best of our knowledge we show for the first time that, in the dipole approximation, where light scattering from realistic particles is identically zero, the atomic SE rate reduces to its value in vacuum even for small distances between the atom and the cloak. This result not only proves that atomic SE decay rate is strongly modified by invisibility of the sphere but also suggests that the emission properties of an atom could be exploited to probe the efficiency of a given plasmonic cloaking device.

This paper is organized as follows. In Sec.~\ref{sec:methods} and ~\ref{sec:plasmonic} we describe the model and the calculation of the atomic SE rate whereas in Sec.~\ref{sec:results}  we present, interpret and discuss our results. Finally, Sec.~\ref{sec:conclusions} is devoted for our conclusions and final remarks.

%--------------------------------------------------------------------------------------------------
\section{The spontaneous emission rate of a two-level system}\label{sec:methods}
In order to establish basic concepts and notation, as well as a convenient expression for the SE rate of an atom in the presence of an arbitrary arrangement of bodies in its surroundings, we start discussing the atom-field dynamics. For simplicity, we consider an atom whose dynamics can be well described by two of its eigenstates. In the absence of interaction we assume that the lowest state $\vert 1 \rangle$ with energy $E_1 = -\hbar\omega_0/2$ has a very long lifetime and a well-defined parity, while the highest-energy state $\vert 2 \rangle$ with energy $E_2 = \hbar\omega_0/2$ has opposite parity and a non-vanishing electric dipole coupling to $\vert 1 \rangle$. Besides, the influence of the aforementioned bodies on the atomic radiative properties are taken into account in this model by the boundary conditions (BC) imposed by them on the quantum electromagnetic field modes. Bearing this in mind, let us assume that the system can be described by the well-known Hamiltonian \cite{Eberly}
\begin{equation}
\hat{{\cal H}} = \hat{{\cal H}}_{\textrm{at}} + \hat{{\cal H}}_{\textrm{f}} + \hat{{\cal H}}_{\textrm{int}} ,
\label{HamiltonianoSistema}
\end{equation}
where $\hat{{\cal H}}_{\textrm{at}}$, $\hat{{\cal H}}_{\textrm{f}}$ and $\hat{{\cal H}}_{\textrm{int}}$ the atomic, the electromagnetic field and the interaction Hamiltonians, respectively. More specifically,
\begin{equation}
\hat{{\cal H}}_{\textrm{at}} = \dfrac{1}{2} \hbar \omega_0 (| 2\rangle \langle 2| - | 1\rangle \langle 1|)
= \dfrac{1}{2} \hbar \omega_0 \hat{\sigma}_{z}
\label{HamiltonianoAtomo}
\end{equation}
is the pure Hamiltonian describing the atomic internal dynamics, $\omega_0$ is the associated transition frequency and $\hat{\sigma}_x, \hat{\sigma}_y, \hat{\sigma}_z$ are the Pauli operators. The second term in (\ref{HamiltonianoSistema}) is given by
\begin{equation}
\hat{{\cal H}}_{\textrm{f}} = \dfrac{1}{8\pi} \int_V(\hat{{\bf E}}^2 + \hat{{\bf B}}^2)d^3 {\bf r},
\label{HamiltonianoCampo}
\end{equation}
where $\hat{{\bf E}}({\bf r})$ and $\hat{{\bf B}}({\bf r})$ are the quantum transverse electric and magnetic field operators. Furthermore, the atom is assumed to be much smaller than all other relevant length scales of the problem, allowing for the electric dipole approximation so that the third term in (\ref{HamiltonianoSistema}) reads
\begin{equation}
\hat{{\cal H}}_{\textrm{int}} = - \hat{{\bf d}}\cdot \hat{{\bf E}},
\label{HamiltonianoInteracao}
\end{equation}
where $\hat{{\bf d}}$ is the atomic electric dipole operator. In the two-level approximation it takes the simple form $\hat{{\bf d}} = {\bf d}_{21}\hat{\sigma}_x$, where ${\bf d}_{21} = \langle2\vert\hat{{\bf d}}\vert1\rangle$ is the transition electric dipole moment that can be made real by an appropriate choice of the relative phase between the ground and the excited states. We also assume the atom has no permanent electric dipole moment so that $\langle1\vert\hat{{\bf d}}\vert1\rangle = \langle2\vert\hat{{\bf d}}\vert2\rangle = {\bf 0}$.

The electromagnetic and interaction terms may be recast in a more convenient form by expanding the electromagnetic field in their normal modes \cite{foot1}
\begin{eqnarray}
\hat{{\bf E}}({\bf r})\!\! &=&	\! i\sum_{\zeta}\sqrt{2\pi\hbar\omega_{\zeta}}
\ [\hat{a}_{\zeta} {\bf A}_{\zeta}({\bf r})-\hat{a}_{\zeta}^{\dagger} {\bf A}_{\zeta}^{*}({\bf r})],\label{campoquantizadogeral0}\\\cr
\hat{{\bf B}}({\bf r})\!\! &=&	 \! \sum_{\zeta} \sqrt{\dfrac{2\pi\hbar}{\omega_{\zeta}}}
\ \nabla \!\times\! [\hat{a}_{\zeta}{\bf A}_{\zeta}({\bf r}) - \hat{a}_{\zeta}^{\dagger}{\bf A}_{\lambda}^*({\bf r})],
\label{campoquantizadogeral}
\end{eqnarray}
where the label $\zeta$ represents an arbitrary complete set of quantum numbers, $\omega_{\zeta}$ are the related eigenfrequencies, and $\hat{a}_{\zeta}, \hat{a}_{\zeta}^{\dagger}$ are the annihilation and creation operators that contain all quantum properties of the field and  satisfy the commutation relations $[\hat{a}_{\zeta},\hat{a}_{\zeta'}^{\dagger}] = \delta_{\zeta\zeta'}$. Note that ${\bf A}_{\zeta}({\bf r})$ are classical functions determined by the Helmholtz equation $ (\nabla^2 + k_{\zeta}^2){\bf A}_{\zeta}({\bf r}) = 0$ and the Coulomb gauge $\nabla \cdot {\bf A}_{\zeta}  ({\bf r})= 0$  with the appropriate BCs.  Substituting Eqs. (\ref{campoquantizadogeral0}) and (\ref{campoquantizadogeral}) into (\ref{HamiltonianoCampo}) and choosing an orthonormal set of functions ${\bf A}_{\zeta}({\bf r})$, namely,
$
%\begin{equation}
\int d^3 r {\bf A}_{\zeta}^{*}({\bf r}) \cdot {\bf A}_{\zeta'}({\bf r}) = \delta_{\zeta \zeta'}
%\label{Ortonormalidade}
%\end{equation}
$,
one can see that up to zero-point contributions that do not contribute to the dynamics, we get
\begin{equation}
\hat{{\cal H}}_{\textrm{f}} = \sum_{\zeta} \hbar\omega_{\zeta}\hat{a}^{\dagger}_{\zeta}\hat{a}_{\zeta}\, .
\label{HamiltonianoCampo2}
\end{equation}
Analogously, we have
\begin{eqnarray}
\hat{{\cal H}}_{\textrm{int}} = -i\hbar\sum_{\zeta}\hat{\sigma}_{x}\left[g_{\zeta}\hat{a}_{\zeta}
 -g^*_{\zeta}\hat{a}_{\zeta}^{\dagger} \right] ,
\label{HamiltonianoInteracao2}
\end{eqnarray}
where we have defined
\begin{equation}
g_{\zeta} \equiv \sqrt{\dfrac{2\pi \omega_{\zeta}}{\hbar}}\; {\bf d}_{21} \cdot
{\bf A}_{\zeta}^{*}({\bf r}) .
\label{gzeta}
\end{equation}

In order to obtain to the SE rate, we have to determine the time evolution of the atomic energy average. Formally, this can be accomplished in the Heisenberg picture by writing an expression for $\hat{\sigma}_z(t)$ and then taking its mean value in the initial state, where the atom is assumed to be in the highest energy level and the field in the vacuum state. Applying the Heisenberg equation, we obtain
\begin{eqnarray}
\dot{{\hat{\sigma}}}^{\dagger}(t) &=& i\omega_0\hat{\sigma}^{\dagger}(t)\! -\!\!\! \sum_{\zeta}
%&&
\!\!\left[g_{\zeta}\hat{\sigma}_{z}(t)\hat{a}_{\zeta}(t) - g_{\zeta}^{*}\hat{a}_{\zeta}^{\dagger}(t)\hat{\sigma}_{z}(t)\!\right]\!\! , \label{EqSigmaDagger} \\
%%%\eea
%
%%%\be
\dot{{\hat{\sigma}}}_z(t) &=& - 2 i  \sum_{\zeta} \left[ g_{\zeta}\hat{\sigma}_y(t)\hat{a}_{\zeta}(t)
- g_{\zeta}^{*}\hat{a}_{\zeta}^{\dagger}(t)\hat{\sigma}_y(t)\right]\! ,\label{EqSigmaZ} \\
%%%\ee
%
%%%\be
\dot{{\hat{a}}}_{\zeta}(t) &=& -i\omega_{\zeta} \hat{a}_{\zeta}(t) + g_{\zeta}\hat{\sigma}_{x}(t) ,
\label{EqOp_a}
%%%\ee
\end{eqnarray}
where we used the normal ordering for operators $\hat{a}_{\zeta}$ and $\hat{a}_{\zeta}^{\dagger}$.
Also, $\hat{\sigma}, \hat{\sigma}^{\dagger}$ are the Pauli lowering and raising operators for the atomic states and are related to $\hat{\sigma}_z$ by the following commutation relations
\begin{eqnarray}
&&[\hat{\sigma},\hat{\sigma}_{z}] = 2 \hat{\sigma} \; , \;
[\hat{\sigma}^{\dagger},\hat{\sigma}_{z}] = -2 \hat{\sigma}^{\dagger}\, , \\
&&\hspace{30pt}\left[ \hat{\sigma},\hat{\sigma}^{\dagger} \right]= -{\hat{\sigma}}_{z} .
\label{CommutationRelations}
\end{eqnarray}

The coupled Eqs. (\ref{EqSigmaDagger}), (\ref{EqSigmaZ}), (\ref{EqOp_a}) are quite difficult to solve, even for the simple system considered here. Besides, a direct integration in Eq. (\ref{EqOp_a}) shows that the relationship between $\hat{a}_\zeta(t)$  and $\hat{\sigma}_x(t)$ is non-local in time, with the field (atomic) operators depending on the atomic (field) operators in earlier times. Fortunately, considering that the atom is weakly coupled to the field, it is possible to simplify considerably these equations by performing a Markovian approximation and arrive at \cite{Milonni, Eberly}
\begin{eqnarray}
\hat{a}_{\zeta}(t) \simeq \hat{a}_{\zeta}^{v}(t) - i g_{\zeta} &&\!\!\!\!\!\!
\left[\hat{\sigma}^{\dagger}(t) \xi^{*}(\omega_{\zeta}+\omega_{0}) \right. \cr\cr
&+& \left. \hat{\sigma}(t)\xi^{*}(\omega_{\zeta}-\omega_{0}) \right],
\label{Solucaoadagger}
\end{eqnarray}
where $\hat{a}_{\zeta}^{v}(t) =  \hat{a}_{\zeta}(0) e^{-i\omega_{\zeta} t}$ is the free field homogeneous solution and
\begin{eqnarray}
\xi(x):= {\cal {P}}\left(\dfrac{1}{x}\right)-\pi i \delta(x),
\label{funcaoxi}
\end{eqnarray}
with ${\cal{P}}$ denoting the principal Cauchy value. Using the last two equations we can rewrite (\ref{EqSigmaZ}) in a simpler way
\begin{eqnarray}
\dot{\hat{\sigma}}_z(t) = &-&2i\sum_{\zeta} g_{\zeta} \left[ \hat{\sigma}_y(t)
\hat{a}_{\zeta}^{v}(t) - \hat{a}_{\zeta}^{v \dagger}(t)\hat{\sigma}_y(t) \right]\cr\cr
&-&2\pi \left[ \mathds{1} + \hat{\sigma}_{z}(t)\right] \sum_{\zeta} |g_{\zeta}|^2
\delta(\omega_{\zeta}-\omega_{0}).
\label{EqSigmaZ2}
\end{eqnarray}
Taking the average value of Eq. (\ref{EqSigmaZ2}) in the initial state $\vert i \rangle = \vert 2 \rangle \otimes \vert\textrm{vacuum} \rangle$ and using $\hat{a}_{\zeta}^{v}(t)\vert\textrm{vacuum} \rangle = 0$, we obtain
\begin{eqnarray}
\langle\hat{\sigma}_{z}(t)\rangle_2 &=& -1 + \left[ \langle\hat{\sigma}_{z}(0)\rangle_2 + 1 \right]
e^{- \Gamma_{21}({\bf r}) t} \nonumber \\
 &=& -1 + 2 e^{- \Gamma_{21}({\bf r}) t},
\label{EqSigmaZ3}
\end{eqnarray}
where $\langle(...)\rangle_2 = \langle2\vert(...)\vert2\rangle$ and we have used that $\langle\hat{\sigma}_{z}(0)\rangle_2 =1$. The most relevant quantity in Eq. (\ref{EqSigmaZ2}) is, however, the spontaneous emission rate $\Gamma_{21}({\bf r})$, given by
\begin{eqnarray}
\Gamma_{21}({\bf r}) = \dfrac{4 \pi^2 \omega_0}{\hbar} \sum_{\zeta} |{\bf d}_{21} \cdot {\bf A}_{\zeta}({\bf r})|^2\delta(\omega_{\zeta}-\omega_{0}) .
\label{TaxaEmissaoEspontanea}
\end{eqnarray}
Equation (\ref{TaxaEmissaoEspontanea}) is the well known result for the SE rate of a two-level atom in the presence of bodies of arbitrary form. The influence of such bodies is coded in the functions ${\bf A}_{\zeta}({\bf r})$ that, as mentioned before, are classical solutions of the Helmholtz equation. This is extremely convenient, as it makes the problem amenable to various
well-developed analytical and numerical techniques. Finally, as an easy but important check, and also for future convenience, one may use Eq.~(\ref{TaxaEmissaoEspontanea}) to obtain the SE rate in free space. In this case, functions ${\bf A}_{\zeta}({\bf r})$ are simply given by
\begin{equation}
{\bf A}_{\zeta}({\bf r}) \, \rightarrow \, {\bf A}_{{\bf k}p}^{(inc)}({\bf r}) = \dfrac{e^{i{\bf k}\cdot {\bf r}}}{(2\pi)^{3/2}} \mbox{{\mathversion{bold}${\epsilon}$}}_{{\bf k} p},
\label{ModesFreeSpace}
\end{equation}
where $\boldsymbol{\epsilon}_{{\bf k} p}$ is a polarization vector. Inserting Eq. (\ref{ModesFreeSpace}) into Eq. (\ref{TaxaEmissaoEspontanea}) we obtain
\begin{eqnarray}
\Gamma_{21}^{(0)} &&= \dfrac{4 \pi^2 \omega_0}{8\pi^3\hbar} |{\bf d}_{21}|^2 \sum_{p=1,2}\int d^3 k \cos^2 \theta
\,\delta(\omega_{\bf k}-\omega_{0}) \nonumber \\
%&&= \dfrac{|{\bf d}_{21}|^2 \omega_{0}}{\hbar} \times 2 \omega_0^2 \int_0^{\pi} d\theta \sin \theta \cos^2 \theta \nonumber \\
&&=\dfrac{4}{3} \dfrac{|{\bf d}_{21}|^2 \omega_{0}^3}{\hbar},
\label{GammaFreeSpace}
\end{eqnarray}
which is the well-known result firstly obtained by P.A.M. Dirac in 1927 \cite{Dirac27}.

%%%%%%%%%%%%%%%%%%%%%%%%%%%%%%%%%%%%%555 %%%%%%%%%%%%%%%%%%%%%%%%%%%%%%%%%%%%%%

\section{Influence of a spherical plasmonic cloak in the spontaneous emission rate} \label{sec:plasmonic}

\begin{figure}[h!]
\centering
\includegraphics[scale=0.43]{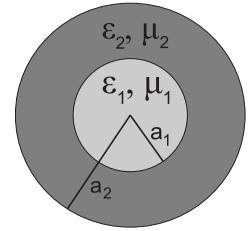}
\caption {Cross section of a our spherical object showing the two radii $a_1$ and $a_2 > a_1$. The inner sphere and
the covering layer are assumed to be made of isotropic homogenous materials, with permittivities and permeabilities $\epsilon_1(\omega), \;\mu_1(\omega)$ and $\epsilon_2(\omega),\;\mu_2(\omega)$, respectively.}
\label{FigSphere}
\end{figure}

In this section we shall calculate the SE rate of an atom near a plasmonic cloak. Following the pioneering work of Alu and Engheta \cite{alu2005jap,alu2005}, the cloak is composed by a sphere with inner radius $a_1$, permittivity $\epsilon_1(\omega)$ covered by a spherical shell with outer radius $a_2 > a_1$ and permittivity $\epsilon_2(\omega)$, as shown in Fig. \ref{FigSphere}. Both the core and the shell are nonmagnetic so that $\mu_1(\omega) = \mu_2(\omega) =  \mu_{0} = 1$.
%
%As can be seen from eq. (\ref{TaxaEmissaoEspontanea}), to calculate the SE rate we just need the electromagnetic field modes expressions explicitly in the presence of the two layered spherical system.

In order to obtain the field modes ${\bf A}_{{\bf k} p}({\bf r})$ we consider the standard Mie approach for the scattering of a single plane wave (polarized in the ${\bf{e}}_1'$ direction and propagating in the ${\bf{e}}_3'$ direction) by the plasmonic sphere. The resulting field at an arbitrary observation point is \cite{BohrenHuffman,Chew}
\begin{eqnarray}
{\bf A}_{{\bf k} p}({\bf r}') = {\bf A}_{{\bf k} p}^{(inc)}({\bf r}') + {\bf A}_{{\bf k} p}^{(sc)}({\bf r}'),
\label{ModosCampo}
\end{eqnarray}
where ${\bf A}_{{\bf k} p}^{(inc)}({\bf r}')$ is given by Eq. (\ref{ModesFreeSpace}) and the scattered contribution can be written as a sum of TM$^r$ and TE$^r$ spherical waves
\begin{eqnarray}
{\bf A}_{{\bf k} p}^{(sc)}({\bf r}')  &=& -\frac{1}{(2\pi)^{3/2}} \left[ \nabla' \times
{\bf F}_{{\bf k}p}^{(sc)}({\bf r}') \right. \nonumber \\
&+& \left.\frac{i}{\omega} \nabla' \times \nabla' \times
{\bf G}_{{\bf k}p}^{(sc)}({\bf r}') \right] \, ,
\label{CampoEspalhado}
\end{eqnarray}
with
\bea
{\bf F}_{{\bf k} p}^{(sc)}({\bf r}') &&= {\bf e}_r'\sum_{l=1}^{\infty} \dfrac{(2l+1)i^{l-1}}{2l(l+1)\omega} c_l^{TE} kr' h_l^{(1)}(kr') \nonumber \\
&&\times \left[ Y_l^1(\theta',\varphi') - Y_l^{-1}(\theta',\varphi') \right] ,\label{FeG0} \\ \cr
{\bf G}_{{\bf k} p}^{(sc)}({\bf r}') &&= {\bf e}_r' \sum_{l=1}^{\infty} \dfrac{(2l+1)i^l}{2l(l+1)\omega} c_l^{TM} kr' h_l^{(1)}(kr') \nonumber \\
&&\times \left[ Y_l^1(\theta',\varphi') + Y_l^{-1}(\theta',\varphi') \right] .
\label{FeG}
\end{eqnarray}
In Eqs.~(\ref{FeG0}) and (\ref{FeG}) $\theta'$ and $\varphi'$ are measured from the ${\bf{e}}_3'$ and ${\bf{e}}_1'$ directions. $Y_l^m(\theta',\varphi')$ are the spherical harmonics \cite{Abramowitz}, and $j_l(x)$ and $h_l^{(1)}(x)$ are the spherical Bessel functions and the spherical Hankel functions of the first kind, respectively \cite{Abramowitz}.  The coefficients $c_l^{TE}$ and  $c_l^{TM}$ are the so-called Mie coefficients \cite{Chew,BohrenHuffman}, and here they are completely determined from the BCs at $r=a_1$ and $r=a_2$. Also, the above expressions have a direct interpretation: the total scattered field is a superposition of the fields produced by all  multipoles induced on the coated sphere.

\begin{figure}[h!]
\centering
\includegraphics[scale=0.3]{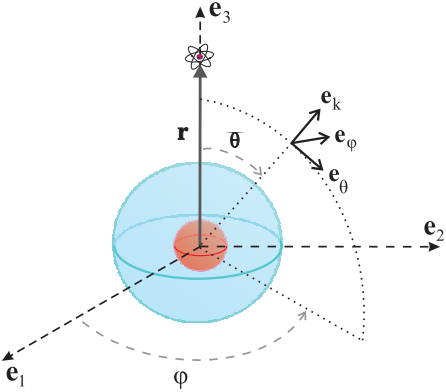}
\caption{(Color online) Coordinate system used throughout the paper. The atom is placed at a distance $r$ from the center of the plasmonic cloak. }
\label{SistemaEixos}
\end{figure}

In order to calculate explicitly the sum in Eq. (\ref{TaxaEmissaoEspontanea}) we now have to rewrite Eqs. (\ref{FeG0}) and (\ref{FeG}) in the unprimed coordinate system shown in the fig. \ref{SistemaEixos}, where the observation point (the position of the atom) lies within the ${\bf{e}}_3$ direction. Formally, this can be accomplished by rotating the spherical harmonics in Eqs. (\ref{FeG0}) and (\ref{FeG}) using the expressions
\bea
 Y_l^{\pm 1}(\theta',\varphi') = \sum_m e^{-im\alpha_k}  d^l_{m, \pm 1} (\beta_k) Y_l^m(\theta,\varphi) ,
\eea
where $d^l_{m, \pm 1} (\theta)$ are the so-called rotation matrices and $\alpha_k$, $\beta_k$ (and also $\gamma_k=0$) are the respective Euler angles. However, in our case this is not necessary since we are rotating the coordinate system in Eqs.~(\ref{CampoEspalhado}), (\ref{FeG0}), and (\ref{FeG}) by precisely the relative angles $\theta'$ and $\varphi'$. Then, by symmetry, it is clear that one just have to evaluate the curls in (\ref{FeG0}) and (\ref{FeG}) and reinterpret the relative angles $\theta'$ and $\phi'$ in terms of the spherical coordinates defined in the Fig. \ref{SistemaEixos} as follows: $r' \rightarrow r$, $\theta' \rightarrow -\theta$ and $\varphi' \rightarrow 0$. Analogously, all the above equations and analysis remain valid in the case in which the incident field is polarized in the ${\bf e}_2'$ direction and propagating in the ${\bf{e}}_3'$ direction; hence it suffices to change the relative angle $\varphi'$ by $\pi/2-\varphi'$.

With those remarks in mind it can be shown that, by retaining only the electric dipole contributions to the scattered fields, the electromagnetic field modes read
%
%\begin{widetext}
\begin{eqnarray}
{\bf A}_{{\bf k} \varphi}^{(sc)}({\bf r}) &&= -\dfrac{3  c_1^{TM}}{2(2\pi)^{3/2}}
\left[\dfrac{h_1^{(1)}(kr)}{kr}
+ {h'}_1^{(1)}(kr) \right] {\bf e}_{\varphi},\label{DipoleFields1}
\end{eqnarray}
\begin{eqnarray}
{\bf A}_{{\bf k}\theta}^{(sc)}({\bf r}) &&=\dfrac{3  c_1^{TM}}{2(2\pi)^{3/2}} \left\{
 \dfrac{\sin 2\theta}{2}
\left({h'}_1^{(1)}(kr) -\dfrac{h_1^{(1)}(kr)}{kr}\right) {\bf e}_{k} \right. \nonumber \\ \cr\cr
\!\!&&\hspace{-1.0cm}+ \left. \left([1+\sin^2\theta]\dfrac{h_1^{(1)}(kr)}{kr}-{h'}_1^{(1)}(kr)\cos^2\theta  \right)
{\bf e}_{\theta} \right\},
\label{DipoleFields2}
\end{eqnarray}
%\end{widetext}
%
where the prime in ${h'}_1^{(1)}(x)$ denotes differentiation with respect to the argument.

It is worth emphasizing that we have two different approximations here: on the one hand, we have the (electric) dipole approximation for the atom, meaning that
$\lambda_0 = 2 \pi c /\omega_0 \gg a_0$, where $a_0$ is the Bohr radius; on the other hand, we have the (electric) dipole approximation for the cloaked sphere, which means that $\lambda_0 = 2 \pi c /\omega_0 \gg a_2$. The former allows one to simplify the interaction part of the Hamiltonian, while the latter justifies the neglect of all the scattering coefficients but $c_1^{TM}$. The dipole approximation for the coated sphere also implies that $|c_1^{TM}| \ll 1$ and therefore  $|{\bf A}_{{\bf k} p}^{(sc)}({\bf r})| \ll |{\bf A}_{{\bf k} p}^{(inc)}({\bf r})|$, so that by substituting Eq. (\ref{ModosCampo}) into Eq. (\ref{TaxaEmissaoEspontanea}) we may keep only linear terms in ${\bf A}_{{\bf k} p}^{(sc)}({\bf r})$, obtaining
\begin{eqnarray}
&&\!\!\!\!\!\!\!\!\!\Gamma_{21}({\bf r}) \simeq \dfrac{4 \pi^2 \omega_0}{\hbar} \sum_{{\bf k}p}  \Bigg\{ |{\bf d}_{21} \cdot
{\bf A}^{(inc)}_{{\bf k}p}({\bf r})|^2 + \label{TaxaEmissaoEspontanea2} \\
\!&+&  2 \, {\cal{R}}e \left( \left[{\bf d}_{21} \cdot
{\bf A}_{{\bf k}p}^{(inc)}({\bf r})\right]^*\!\! \left[{\bf d}_{21}
\cdot{\bf A}_{{\bf k}p}^{(sc)}({\bf r}) \right] \right)\Bigg\} \delta(\omega_k-\omega_0).\nonumber
\end{eqnarray}

The first term in the previous equation gives the free space contribution (\ref{GammaFreeSpace}), while the second one is precisely the correction due to the presence of the coated sphere. A direct substitution of Eqs. (\ref{DipoleFields1}) and (\ref{DipoleFields2}) into Eq. (\ref{TaxaEmissaoEspontanea2}) yields, after a lengthy but straightforward calculation,
\begin{widetext}
\begin{eqnarray}
\label{emissionfinal}
\dfrac{\Delta\Gamma_{21}({\bf r})}{\Gamma_{21}^{(0)}}\simeq \dfrac{3}{2}|c_1^{TM}|
\left\{\dfrac{n_1(k_{0}r)}{k_{0}r}\left[-\dfrac{\sin(k_{0}r)}{(k_{0}r)^3}+\dfrac{2\cos(k_{0}r)}{(k_{0}r)^2}-
 \dfrac{\sin(k_{0}r)}{k_{0}r} \right]\right.
 + \left. 2{n'}_1(k_{0}r)\left[-\dfrac{\sin(k_{0}r)}{(k_{0}r)^3}+
\dfrac{\cos(k_{0}r)}{(k_{0}r)^2}
+ \dfrac{\sin(k_{0}r)}{k_{0}r} \right]\right\}\cr
\phantom{oi}
\end{eqnarray}
\end{widetext}
where $\Delta\Gamma_{21}({\bf r}) = \Gamma_{21}({\bf r}) -  \Gamma_{21}^{(0)}$, $k_0 = 2\pi/\lambda_0$ and $n_1(x)$ is the spherical Neumann function~\cite{Abramowitz}. In order to obtain the previous result we considered an isotropic atom, $|{d_{21}}_x|^2 = |{d_{21}}_y|^2 = |{d_{21}}_z|^2 = |{\bf d}_{21}|^2/3$, and neglected losses by assuming real permittivities at the transition frequency.

Equation (\ref{emissionfinal}) is the central result of this section and it demonstrates that SE rate of a two-level atom placed in the vicinity of a spherical plasmonic cloak is proportional to the first TM harmonic of the scattering coefficient of the coated sphere, $|c_1^{TM}|$. As shown by Al\`u and Engheta, this coefficient, which corresponds to the electric dipole radiation, can be made to vanish by a judicious choice of material parameters~\cite{alu2005jap,alu2005,alu2008}. As a consequence, the scattering cross section corresponding to the coated sphere will be greatly reduced since $|c_1^{TM}|$ largely dominates the scattering pattern, making the system practically invisible to the incident radiation. This result has an interesting impact on the atomic SE rate, as it will be discussed in the next section.

%
%--------------------------------------------------------------------------------------------------
%
\section{Discussions}
\label{sec:results}

In the electric dipole approximation, the coefficient $c_1^{TM}$ reads
\begin{equation}
c_1^{ TM} \simeq i \dfrac{U_1^{TM}}{V_1^{TM}}\, ,
\end{equation}
where in the limit of low losses $U_1^{TM}$ and $V_1^{TM}$ are real functions and can be written as
\begin{eqnarray}
U_1^{TM} &\simeq& \dfrac{(k a_2)^2}{3}\left|
\begin{array}{cccc}
1              &              1            &             -1        &0 \\
2\varepsilon_1^{-1} & 2\varepsilon_2^{-1}            &\varepsilon_2^{-1}        &0  \\
0              &\gamma^{-1}                &-\gamma^2              &1 \\
0              &2(\gamma\varepsilon_2)^{-1} & \gamma^2 \varepsilon_2^{-1} & 2\varepsilon_2^{-1}\\
\end{array}
\right| \, , \\ \cr\cr
V_1^{TM} &\simeq& \dfrac{1}{k a_2}\left|
\begin{array}{cccc}
1              &              1            &             -1        &0 \\
\varepsilon_1^{-1} & \varepsilon_2^{-1}            &\varepsilon_2^{-1}        &0 \\
0              &\gamma^{-1}                &-\gamma^2              &-1 \\
0              &2(\gamma\varepsilon_2)^{-1} & \gamma^2 \varepsilon_2^{-1} & \varepsilon_2^{-1} \\
\end{array}
\right|\, ,
\end{eqnarray}
with $\gamma = a_1/a_2$ and we made the vacuum electric permittivity equal to one, $\varepsilon_0=1$.

The effectiveness of a cloak is related to the amount of scattering that is suppresed. In particular, an ideal cloak is characterized by no scattering at all, i.e., by the vanishing of $c_1^{ TM}$. It can be shown that a sufficient condition for that to happen is~\cite{alu2005,alu2008},
\begin{equation}
\label{condition}
\gamma = \dfrac{a_1}{a_2} = \sqrt[3]{\dfrac{(\varepsilon_2-1)(2\varepsilon_2+\varepsilon_1)}{(\varepsilon_2-\varepsilon_1)(2\varepsilon_2+1)}}.
\end{equation}
The complete set of material parameters for which invisibility can occur has been carefully discussed in Refs.~\cite{alu2005jap,alu2005,alu2008}, and here we explore the case where $\epsilon_{i,j}(\omega_0) < 1 < \epsilon_{j,i}(\omega_0)$, with $i,j=1,2$.

\begin{figure}[h!]
\centering
\includegraphics[scale=0.42]{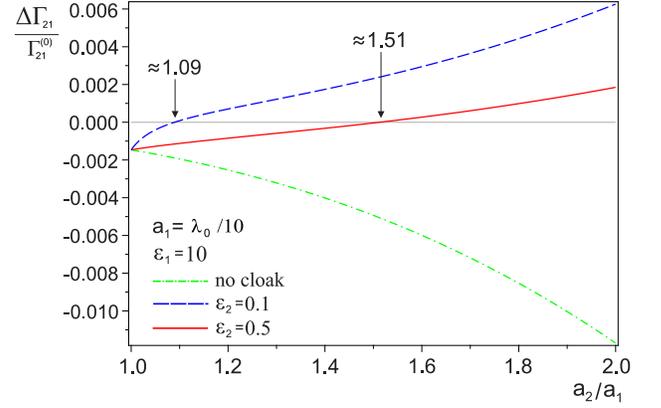}
\caption{(Color online) Relative spontaneous emission rate $\Delta\Gamma_{21}({\bf r})/\Gamma_{21}^{(0)}$ as a function of $a_2/a_1$ for a given distance between the sphere and the atom. The sphere parameters have the following fixed values: $a_1 = \lambda_{0}/10$ and $\varepsilon_1 = 10$. The blue dashed line and the red solid line show the results for $\varepsilon_2 = 0.1$ and $\varepsilon_2 = 0.5$,  respectively. Vertical arrows indicate the values of $a_2/a_1$ for which the SE rate reduces to its value in the vacuum. For comparison, the green dotted-dashed line shows the result for the case where the coavering layer is made of the same material as the inner sphere (no cloak).}
\label{EmissaoEspontaneaVersusRaioCloak}
\end{figure}

In Figure \ref{EmissaoEspontaneaVersusRaioCloak} we show $\Delta\Gamma_{21}({\bf r})/\Gamma_{21}^{(0)}$, the difference between the SE rate of a two-level atom near the coated sphere and its vacuum value normalized by $\Gamma_{21}^{(0)}$, as a function of $a_2/a_1$ for a given distance between the atom and the sphere. We fixed the physical parameters characterizing the core by setting the electric permittivity of the inner sphere as $\varepsilon_1 = 10$ and, in order to be consistent with the dipole approximation, we chose the radius $a_1 = \lambda_0/10$. For comparison, in Fig.\ref{EmissaoEspontaneaVersusRaioCloak} also it is also plotted the result for $\Delta\Gamma_{21}({\bf r})/\Gamma_{21}^{(0)}$ in the case where there is no cloak at all (green dot-dashed line), i.e., the relative SE rate in the presence of a single sphere with $\varepsilon_1=10$ as a function of its radius. In this case, as the radius of the sphere grows the SE rate deviates increasingly from its value in vacuum $\Gamma_{21}^{(0)}$, as expected. In contrast, when the shell is taken into account, $\Delta\Gamma_{21}({\bf r})$ approaches zero, and hence $\Gamma_{21}({\bf r})\rightarrow \Gamma_{21}^{(0)}$,  as the radius $a_2$ of the cloak is increased. More interestingly, the SE rate  $\Gamma_{21}({\bf r})$ is identical to its value in vacuum for $a_2/a_1 \simeq 1.09$ (when $\epsilon_2 = 0.1$) and for $a_2/a_1 \simeq 1.51$ (when $\epsilon_2 = 0.5$),  precisely the cases where condition (\ref{condition}) is satisfied for the respective choices of $\epsilon_2$. As far as we know, this is the first situation where the SE rate of a quantum emitter is unaffected by the presence of a surrounding body for a broad range of separation distances. For values of $a_2/a_1$ larger than the condition of invisibility, the contribution of the shell to the scattered field is dominant and the system becomes visible, so that the relative SE rate increases again as can be seen from Fig.~\ref{EmissaoEspontaneaVersusRaioCloak}.

\begin{figure}[h!]
\centering
\includegraphics[scale=0.44]{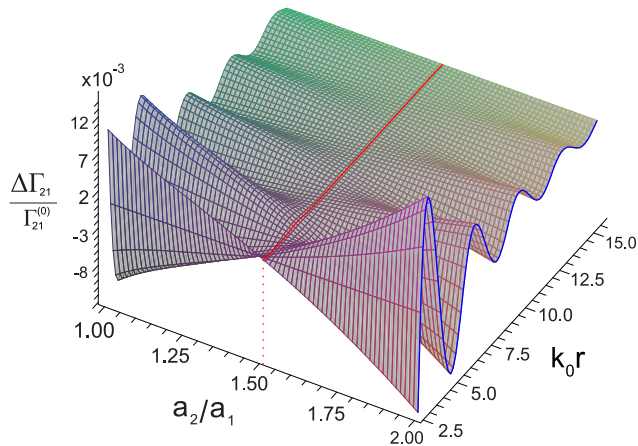}
\caption{(Color online) Relative spontaneous emission rate as function of $a_2/a_1$ and $k_{0} r$. Parameters of the inner sphere are $a_1=\lambda_0/10$ and $\varepsilon_1=10$, while the covering layer has $\varepsilon_2=0.5$. The red solid line highlights the value $a_2/a_1=\sqrt[3]{38/11}\simeq 1.51$, for which the invisibility condition (\ref{condition}) is fulfilled. We note also that $a_2/a_1 = 1$ corresponds to the SE rate of an atom in the presence of a dielectric sphere without cloak. }
\label{EmissaoEspontanea3D}
\end{figure}

In order to investigate the dependence of the SE rate on the distance between the atom and the spherical plasmonic cloak, in Fig. \ref{EmissaoEspontanea3D} we exhibit a three-dimensional plot of $\Delta\Gamma_{21}({\bf r})/\Gamma_{21}^{(0)}$ as a function of both $a_2/a_1$ and $k_{0} r$. The parameters characterizing the inner sphere are the same as in Fig.~\ref{EmissaoEspontaneaVersusRaioCloak} and the electric permittivity of the shell is $\varepsilon_2=0.5$. It is important to emphasize that the SE rate exhibits by and large an oscillatory behavior with the distance $r$ except for $a_2/a_1 = \sqrt[3]{38/11} \simeq 1.51$ (highlighted by the red line in Fig. \ref{EmissaoEspontanea3D}), where the invisibility condition (\ref{condition}) is fulfilled and the SE rate is always equal to $\Gamma_{21}^{(0)}$ for any value of $k_0r$.

\begin{figure}
\centering
\includegraphics[scale=0.4]{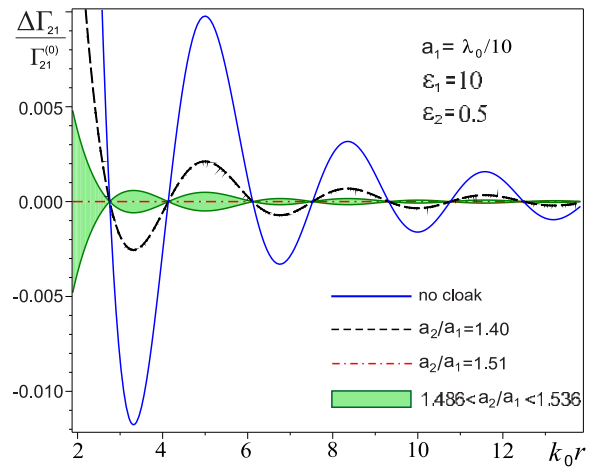}
\caption{(Color online) Relative SE rate  as a function of $k_{0} r$ for different values of  $a_2/a_1$. The blue solid line correspponds to the absence of a cloaking shell, namely, $a_2/a_1=1$. The black dashed line corresponds to $a_2/a_1 = 1.40$ and shows that the covering layer significantly reduces the amplitude of oscillation. The red dotted-dashed line shows the case where the condition (\ref{condition}) is fullfilled, $a_2/a_1 = \sqrt[3]{38/11} \simeq 1.51$. The shaded area corresponds to the interval $1.486<a_2/a_1<1.536$ and highlights the range of values of $a_2/a_1$ for which $\Delta\Gamma_{21}({\bf r})/\Gamma_{21}^{(0)}$ is reduced to 5\% or less than its value in the presence of a single sphere.}
\label{EmissaoEspontaneaVersusDistancia}
\end{figure}

To further explore the dependence of the SE rate upon the shell radius, in Fig.  ~\ref{EmissaoEspontaneaVersusDistancia} we plot $\Delta\Gamma_{21}({\bf r})/\Gamma_{21}^{(0)}$ for different values of the ratio $a_2/a_1$; all other parameters being the same as those used in Fig.~\ref{EmissaoEspontanea3D}. Again, it is clearly seen that the presence of the covering layer significantly reduces the amplitude of oscillation with respect to SE rate for a single sphere. The black dashed line corresponds to $a_2/a_1 = 1.40$ and demonstrates that a reduction of about 80\% for $\Delta\Gamma_{21}({\bf r})/\Gamma_{21}^{(0)}$ (compared to the case without the shell) can be obtained even with plasmonic cloaks for which the invisibility condition (\ref{condition}) is not strictly satisfied. The green shaded area highlights the interval of possible values of $a_2/a_1$ for which the relative SE rate is reduced  by at least 95\% due to the inclusion of the shell. More specifically, the results show that when $1.486\leq a_2/a_1 \leq 1.536$, then $\Gamma_{21}({\bf r})/\Gamma_{21}^{(0)}$ is smaller than 5\% of its bare value, regardless of the distance between the atom and the center of the cloak. This shows that a tolerance of 5\% in the efficiency of the plasmonic cloak is possible in a relatively wide range of ratios $a_2/a_1$, highlighting a serendipitous robustness of our results.

Let us now consider some realistic parameters in order to assess the experimental viability to test the results here discussed.
A well-suited quantum emitter is a Rubidium atom prepared in a Rydberg state with principal number $n=51$ and magnetic number $m=50$, since it is possible to show that such an atom can be well described by a two-level system and therefore our previous discussion can be applied \cite{Brune1994}. Moreover, the corresponding transition frequency is $\nu_0 = 2\pi\omega_0 \approx 51.099$ GHz, well in the range of recent experiments with plasmonic cloaking devices~\cite{edwards2009,Alu2012}. Suppose the Rubidium atom is placed near a spherical cloak composed of a non-magnetic sphere of radius $a_1$ and permittivity $\varepsilon_1 = 10$  covered by a spherical shell with outer radius $a_2$ and permittivity $\varepsilon_2 = 0.5$, being both permittivities evaluated at the transition frequency $\omega_0$. As before, let us take $a_1=\lambda_0/10=587.1\mu$m. Therefore, the ideal value for $a_2$ for a perfect cloaking is $a_2=\sqrt[3]{38/11}\times587.1 \simeq 887.5\mu$m. If, instead of perfect cloaking, we require only that the SE rate is suppressed by at least $95\%$,  the possible values of $a_2$ are in the interval $872.4\mu m\leq a_2\leq901.8\mu m$. In other words, one could vary $a_2$ by $\pm 15\mu$m around the ideal value and would still have a very efficient reduction of the SE rate ($\ge 95\%$). Had we been interested in a $99\%$ of the SE rate, the allowed range for $a_2$ would be narrower, namely, $ 884.8\mu m\leq a_2\leq 890.7\mu m$ (in this case, $a_2$ could be changed by $\pm 3\mu$m around the ideal value). These numbers show that, at least in principle, our result could be tested with current apparatuses and techniques.

% -------------------------------------------------------------------------------------------------
\section{Conclusions}\label{sec:conclusions}

We have investigated the SE rate of a two-level atom placed in the vicinity of a plasmonic cloak composed of a coated sphere. In the dipole approximation, i.e., when the coated sphere is much smaller than the atomic transition wavelength, we have demonstrated that the difference between the SE rate in the presence of the cloak and its vacuum value is directly proportional to the first TM scattering coefficient. Since this coefficient can vanish under certain circumstances, as established by Al\'u and Engheta~\cite{alu2005,alu2005}, this result implies that the SE rate of a two-level atom can be identically reduced to its vacuum value. We have analysed the dependence of the SE rate on the distance between the atom and the plasmonic cloak; we conclude that the reduction of the SE rate is significative for a large range of distances, even for small distances between the atom and the cloak. We have also investigated the dependence of the SE rate on the geometrical parameters of the cloak, such as the ratio between its inner and outer radii, as well as on its material parameters, such as the electrical permittivities; we find that the strong suppression of the SE rate is robust against the variation of both the geometrical and material parameters of the cloak, as a manifestation of the robustness of the plasmonic cloaking mechanism itself.

In order to envisage the possible experimental verification of our findings, we have made realistic estimates for the SE rate of atoms in the vicinities of a plasmonic cloak. Since the frequency operation range of state-of-the-art plasmonic cloaks is typically of a few GHz~\cite{edwards2009,Alu2012}, we argue that the presented results could be experimentally verified as there are many atomic species that spontaneously emit in this frequency range, such as Rubidium, the explicit example we have considered. As a result, we suggest that the observation of the reduction of the atomic SE rate to its vacuum value in the presence of plasmonic cloaks could be explored as an alternative, quantum probe of the effectiveness of these devices.

\section{Acknowledgments}

We would like to thank R. M. Souza and M. V. Cougo-Pinto for stimulating discussions. We also acknowledge CNPq, CAPES and FAPERJ for partial financial support.

% -------------------------------------------------------------------------------------------------


\begin{thebibliography}{99}


\bibitem{Purcell-46} E. M. Purcell,
Phys. Rev. \textbf{69}, 681 (1946).

\bibitem{DrexhageEtAl-68} K.H. Drexhage, H. Kuhn and F.P. Sh\"afer,
 Phys. Chem. {\bf 72}, 329 (1968).

\bibitem{Morawitz-69} H. Morawitz,
 Phys. Rev. {bf 187}, 1792 (1969).

\bibitem{Drexhage-70} K.H. Drexhage,
 Sci. Am. {\bf 222}, 108 (1970).

\bibitem{Drexhage-74} K.H. Drexhage,
in {\it Progress in Optics}, ed. E. Wolf (North-Holland, Amstrdam, 1974), Volume {\bf 12}.

\bibitem{Barton-70-87} G. Barton, Proc. Roy. Soc. Lond. {\bf A320}, 251 (1970); {\bf A410}, 141 (1987).

\bibitem{Stehle-70} P. Stehle, Phys. Rev. A{\bf 2}, 102 (1970).

\bibitem{Milonni-Knight-73} P.W. Milonni and P.L. Knight, Opt. Commun. {\bf 9}, 119 (1973).

\bibitem{Philpott-73} M.R. Philpott, Chem. Phys. Lett. {\bf 19}, 435 (1973).

\bibitem{Alves-Farina-Tort-2000} D.T. Alves, C. Farina and A.C. Tort, Phys. Rev. A{\bf 61}, 34102 (2000).

\bibitem{MendesEtAl-08} T.N.C. Mendes, F.S.S. Rosa, A. Ten\ 'orio and C. Farina,
 Phys. Rev. A{\bf 78}, 012105 (2008).

\bibitem{blanco2004} L. A. Blanco and F. J. Garc\'ia de Abajo,
\prb \textbf{69}, 205414 (2004).

\bibitem{thomas2004} M. Thomas, J.-J. Greffet, R. Carminati, and J. R. Arias-Gonzalez,
\apl \textbf{85}, 3863 (2004).

\bibitem{carminati2006} R. Carminati, J.-J. Greffet, C. Henkel, and J. M. Vigoureaux,
Opt. Commun. \textbf{261}, 368 (2006).

\bibitem{vladimirova2012} Y. V. Vladimirova, V. V. Klimov, V. M. Pastukhov, and V. N. Zadkov,
\pra \textbf{85}, 053408 (2012).

\bibitem{klimovreview} V. V. Klimov, M. Ducloy, and V. S. Letokhov,
Sov. J. Quantum Electron. \textbf{31}, 569 (2001).

\bibitem{Hulet-Hilfer-Kleppner-85} R.G. Hulet, E.S. Hilfer and D. Kleppner,
Phys. Rev. Lett. {\bf 55},2137 (1985).

\bibitem{JheEtAl-87} W. Jhe, A. Anderson, E.A. Hinds, D.Meschede, L. Moi and S. Haroche,
Phys. Rev. Lett. {\bf 58}, 666 (1987).

\bibitem{DeMartiniEtAl-87} F. DeMartini, G. Innocenti, G.R. Jacobovitz and P. Mataloni,
Phys. Rev. Lett. {\bf 59}, 2955 (1987)

\bibitem{HeinzenEtAl-87} D.J. Heinzen, J.J. Childs, J.F.Thomas and M.S. Feld,
Phys. Rev. Lett. {\bf 58}, 1320 (1987).
\bibitem{kuhnreview} S. Kuhn, G. Mori, M. Agio, and V. Sandoghdar,
Mol. Phys. \textbf{106}, 893 (2008).

\bibitem{Milonni} P.W. Milonni, {\sl The Quantum Vacuum. An Introduction to Quantum Electrodynamics} (Academic, San Diego, 1994).

\bibitem{Haroche-Kleppner-89} S. Haroche and D. Kleppner,
{\it Cavity Quantum electrodynamics}, Physics Today (January 1989), 25.

\bibitem{Hinds-90} E.A. Hinds, {\it Cavity Quantum Electrodynamics}, in Advances in Atomic, Molecular and Optical Physcs, ed. D.R.Bates and B. Bederson (Academic Press, Boston 1990), Volume 28.

\bibitem{Berman-94} {\it Cavity Quantum Electrodynamics}, ed. P. Berman (Academic, New York, 1994).

\bibitem{betzig1993} E. Betzig, R. J. Chichester,
Science \textbf{262}, 5138 (1993).

\bibitem{bian1995} R. X. Bian, R. C. Dunn, X. S. Dunn, and P. T. Leung,
\prl \textbf{75}, 4772 (1995).

\bibitem{sanchez1999} E. J. Sanchez, L. Novotny, and X S. Xie,
\prl \textbf{82}, 4014 (1999).

\bibitem{greffet2005} J.-J. Greffet,
Science \textbf{308}, 1561 (2005).

\bibitem{muhlschlegel2005} P. Muhlschlegel, H. J. Eisler, O. J. F. Martin, B. Hecht, D. W. Pohl,
Science \textbf{308}, 1607 (2005).

\bibitem{jackson2004} J. B. Jackson and N. J. Halas,
Proc. Natl. Acad. Sci. USA \textbf{101}, 17930 (2004).

\bibitem{wei2008} H. Wei, F. Hao, Y. Huang, W. Wang, P. Nordlander, and H. Xu,
Nano Lett. \textbf{8}, 2497 (2008).

\bibitem{li2010} J. F. Li {\it et al,},
Nature (London) \textbf{464}, 392 (2010).

\bibitem{klimov2012} Y. V. Vladimirova, V. V. Klimov, V. M. Pastukhov, and V. N. Zadkov,
Phys. Rev. A \textbf{85} 053408 (2012).

\bibitem{Smith00}
%Composite Medium with Simultaneously Negative Permeability and Permittivity
D. R. Smith, W. J. Padilla, D. C. Vier, S. C. Nemat-Nasser, and S. Schultz,
\prl {\bf 84}, 4184 (2000).

\bibitem{smith2004}
%Metamaterials and Negative Refractive Index
D. R. Smith, J. B. Pendry, and M. C. K. Wiltshire,
Science \textbf{305}, 788 (2004).

\bibitem{enkrich2005} C. Enkrich, M. Wegener, S. Linden, S. Burger, L. Zschiedrich, F. Schmidt, J. F. Zhou, Th. Koschny, and C. M. Soukoulis,
Phys. Rev. Lett. \textbf{95}, 203901 (2005).

\bibitem{cai2007} W. Cai, U. K. Chettiar, H. K. Yuan, V. C. de Silva, A. V. Kildishev, V. P. Drachev, and V. M. Shalaev,
Opt. Express \textbf{15}, 3333 (2007).

\bibitem{pendry2006} J. B. Pendry, D. Shurig, and D. R. Smith,
Science \textbf{312}, 1780 (2006).

\bibitem{leonhardt2006} U. Leonhardt, Science \textbf{312}, 1777 (2006).

\bibitem{OPN} N. Zheludev and N. Papasimakis, in \textit{Metamaterial-Induced Transparency: Sharp Fano Resonances and Slow Light}, Optics and Photonics News, \textbf{20} $N^{0}$ 10 (2009).

\bibitem{jacob2012} Z. Jacob, I. I. Smolyaninov, and E. E. Narimanov, Appl. Phys. Lett. \textbf{100}, 181105 (2012).

\bibitem{klimov2002} V. V. Klimov, Opt. Commun. \textbf{211}, 183 (2002).

\bibitem{hyperbolicreview} C.L. Cortes, W. Newman, S. Molesky, and Z. Jacob, J. Opt. \textbf{14}, 063001 (2012).

\bibitem{alu2005jap} A. Al\`u and N. Engheta, J. Appl. Phys. \textbf{97}, 094310 (2005).

\bibitem{alu2005} A. Al\`u and N. Engheta, \pre \textbf{72}, 016623 (2005).

\bibitem{alu2008} A. Al\`u and N. Engheta, J. Opt. A \textbf{10}, 093002 (2008).

\bibitem{edwards2009} B. Edwards, A. Al\`u, M. G. Silveirinha, and N. Engheta, \prl \textbf{103}, 153901 (2009).

\bibitem{Eberly} L. Allen and J. H. Eberly, \textit{Optical ressonance and two-level atoms} (John Wiley \& Sons, New York, 1975).

\bibitem{foot1} From now on we assume that the atom is placed in vacuum, so that
an expansion like (\ref{campoquantizadogeral}) is always possible in the vicinities of the atom.

\bibitem{Dirac27} P.A.M. Dirac, Proc. Roy. Soc. Lond. {\bf 114}, 243 (1927).

\bibitem{Chew} W. C. Chew, {\it Waves and Fields in Inhomogeneous Media}, IEEE Press (1995).

\bibitem{BohrenHuffman} Craig F. Bohren, Donald R. Huffman, {\it Absorption and Scattering of Light by Small
Particles}, John Wiley and sons, Inc. (1983).

\bibitem{Abramowitz}  M. Abramowitz and I.A. Stegun (eds.), {\it Handbook of Mathematical Functions},
(Dover, New York) (1964).

\bibitem{Brune1994} M. Brune, P. Nussenzveig, F. Schmidt-Kaler, F. Bernardot, A. Maali, J. M. Raimond, and S. Haroche, \prl \textbf{72}, 3339 (1994).

\bibitem{Alu2012} D. Rainwater, A. Kerkhoff, K. Melin, J. C. Soric, G. Moreno and A. Al\'u, New J. Phys. \textbf{14}, 013054 (2012).

\end{thebibliography}
\end{document}